\documentclass[24pt]{article}
\usepackage{latexsym}
\usepackage{amsmath,amscd,amsfonts,amssymb,graphicx}
\author{A. V. B. Cruz$^a$, A. K. Mishra$^a$ 
and W. Schmickler$^b$ \\
$^a$ Institute of Mathematical Sciences\\CIT Campus, Chennai, 600113 India\\
$^b$ Institute of Theoretical Chemistry, Ulm University\\
D89069 Ulm, Germany}
\title{Electron Transfer rate between a electrode and a bridged redox}
\begin{document}
\maketitle
{\small We derive an explict bias dependent expression for electron transfer reaction rate from a solvated redox to a electrode through a bridged molecule of arbitrary length. The interaction of the solvated redox with the solvent is modelled as a classical harmonic oscillator bath. The effect of competing process, namely resonance tunneling between redox and bridge and the solvation of the redox is investigated. Plots were produced for the case of 5 atom bridge. Our analysis shows that for certain suitable value of nearest-neighbour coupling, it is possible to block electron transfer in certain intermediate voltage regime}   
\begin{abstract}

\end{abstract}

\section{Introduction}

\vspace{0.2in}
\hspace{0.2in}

{\small Electron transfer at molecular levels and ,especially, between chemically active species has been an important and interesting field of active research for a couple of decades \cite{Kuznetsov01,Kuznetsov02,Schmickler01,Schmickler02}. A typical quantity of interest in chemistry is the rate of transfer between two chemiactive species, that is, between donor and acceptor species which in general are solvated. Additionally, with the advent of nanotechnology and molecular electronics, understanding electron transfer rate in molecular chains became a key theoretical interest. A general setup in molecular electron transfer is to study, both in theory as well as in actual experiment, the current-voltage response ontained on passing current between two molecular mesoscopic junctions connected by a single long molecule or a chain of repeated molecular or atomic units. The important quantity of interest in the above scenario is the molecular conduction.  The problem of obtaining the conduction of a such a system is well studied one. Formal expressions and relationships like the landauer formula \cite{Landauer01,Buttiker01,Buttiker02} and it's several variants are available to calculate the conductance of such systems \cite{Imry01,Caroli01,Datta01,Amato01}.  Since conductance is due to transfer of electrons, the question of relationship between conductance and electron transfer rate was answered by A. Nitzan and co-workers \cite{Nitzan01,Nitzan02}. Nitzan derived a relationship between electron transfer rate ($\kappa$) and conductance at certain regime. }

\vspace{0.2in}
\hspace{0.2in}

{\small The rate expression for electron transfer has been derived by several methods, the earliest dating back to the superexchange method proposed initially by McConnell \cite{Connell01,Evenson01,Evenson02,Marcus01,Beratan01,Stuchebrukhov01,Goldman01}. Superexchange treatment is a viable method when the transfer mechanism is tunneling dominated. Another plausible mechanism for transfer is sequential hopping. Typical cases of sequential hopping occurs when there are asymmetries or irregularties in the bridge connecting the acceptor and the donor. Under those conditions phase loss can lead to sequential hopping. The interplay between the Sequential hopping and tunneling has been adressed by Medvedev and Stuchebrukhov \cite{Medvedev01}. They obtained the expression for rate by forming a dynamic correlation function of the couplings or the hopping parameter. Alternate methods to arrive at the rate expression include, density matrix formulation of the problem and equation of motion for the reduced quantum system interacting \cite{Felts01,Felts02,Mukamel01,Mukamel02,Tanimura01,Tanimura02}. } 

\vspace{0.2in}
\hspace{0.2in}

{\small Our present work focuses on obtaining an explict expression for transfer rate in a system where the redox is solvated and is connected by a molecular chain to a electrode. More specifically our attention is concentrated towards obtaining the voltage dependecy of the electron transfer rate. As with the previous authors who tackled the problem within the time-dependent fluctuation framework, we assume that at t=0 the electron is in the donor and express the rate of the electron arrival at the final continuum states in the electrode using t-matrix. In this way our treatment of the problem is different from the previous treatments by other authors, wherein the rate expression was derived by expressing the rate as a time dependent correlation function and by partial tracing of the density matrix.  It should also be noted that all the earlier works for obtained transfer rate were mainly oriented towards a DBA system or some likewise where both the donor and acceptor were either solvated or continuum of states, while in our system considered only the donor is solvated while the acceptor is an electrode. }

\vspace{0.2in}
\hspace{0.2in}

\section{Model and Calculation}

\vspace{0.2in}
\hspace{0.2in}

{\small The following Hamiltonian is considered as a model for a system of N-sites (single energy level $\epsilon_{i}$)  connecting a reservoir or electrode with energy levels $\epsilon_{k}$ to a redox couple with energy level $\epsilon_{r}$. The electrode and the redox is connected to the N-site chain through the 1 and N- site.  The Hamiltonian is a tight binding model hamiltonian with additional interactions to account for the interaction of the adsorbate and the redox with the polrization modes. This is acheieved by treating these polarization modes as classical phonon modes. For ease of clarity the Hamiltonian can be split into three parts $H_{el}$, $H_{chain}$ and  $H_{ph}$, where the $H_{el}$ refers to the electrode part of the Hamiltonian  , $H_{chain}$ represents the tight-binding Hamiltonian of the chain + redox couple and the $H_{ph}$ is the used for modelling the classical phonon modes. }

\begin{equation}
\mathrm{H}_{el} = \Sigma_{k} \epsilon_{k} n_{k} + \Sigma_{k} \lbrack \bar{\upsilon}_{k1} c^{\dagger}_{k} c_{1} + h.c \rbrack 
\end{equation}

\begin{equation}
\mathrm{H}_{ph} = \Sigma_{\nu} \frac{1}{2} \hbar \omega_{\nu} q_{\nu} ^{2}
\end{equation}

\begin{equation}
 \mathrm{H}_{chain} = \sum_{i=1,N} \epsilon_{i} n_{i} + \sum_{i} \lbrace \upsilon_{i} c_{i}^{\dagger}c_{i+1} + h.c \rbrace + (\epsilon_{r} + \sum_{\nu} \hbar \omega_{\nu} g_{\nu} q_{\nu} )n_{r} + \lbrace \bar{\upsilon}_{nr} c_{n}^{\dagger} c_{r} + h.c \rbrace
\end{equation}

\vspace{0.2in}
\hspace{0.2in}

{\small It is clear that the matrix form for the $\mathrm{H}_{chain}$ is a N+1 $\times$ N+1 tridiagonal matrix with only non-zero entries along the main diagonal and sub-diagonals. For completeness we also give below the matrix form of the hamiltonian}

\begin{equation}
\mathrm{H}_{chain} = \left( \begin{array}{ccccc}
\epsilon_{1}  & \upsilon & \ldots & 0 & 0  \\
\upsilon & \epsilon_{2} & \upsilon & \ldots  & 0\\
\ldots & \ldots & \ldots & \upsilon & 0\\
0 & \ldots & \upsilon & \epsilon_{N} & \bar{\upsilon}_{rn}\\
0 & \ldots & \ldots &  \bar{\upsilon}_{nr} & \epsilon_{r} + \sum_{\nu} \hbar \omega_{\nu} g_{\nu} q_{\nu} \\
\end{array} \right)
\end{equation}

\begin{equation}
\mathrm{H} = \mathrm{H}_{el} + \mathrm{H}_{chain} + \mathrm{H}_{ph}
\end{equation}

\vspace{0.2in}
\hspace{0.2in}
{\small To get the current the formalism of Ratner et al is employed. Thus the procedure is to get the rate for the transition from redox (r) to the electrode (k) under the influence of voltage W and then perform a thermal averaging of the transition rate. }

\begin{equation}
\mathrm{rate} = \frac{2 \pi}{\hbar} \Sigma_{k} \lbrack 1 - f(\epsilon - eW) \rbrack \mid \mathrm{T}_{kr} \mid ^{2} \delta(\epsilon_{r} + \Sigma_{\nu} \hbar g_{\nu} q_{\nu} - \epsilon_{k})
\end{equation}

\begin{equation}
\mathrm{T}_{kr} = \bar{\upsilon}_{k1} \mathrm{G}_{1n} \bar{\upsilon}_{nr} 
\end{equation}

\begin{equation}
\mathrm{G}_{1n}^{0} = \frac{(-v)^{n-1}}{\mathrm{det} \lbrack \tilde{\mathrm{H}}_{chain} \rbrack _{1n}}
\end{equation}

\vspace{0.2in}
\hspace{0.2in}

{\small The $\tilde{\mathrm{H}}_{chain}$ refers to the Hamiltonian of the chain along with the self-energy correction arising from the coupling of it's 1-st site with the electrode. In effect, this amounts to adding the self-energy to the 1,1 element of the $\mathrm{H}_{chain}$. The $\mathrm{G}_{1n}^{0}$ is calculated based on the following crucial observation.  Let $d_{n}$ denote the determinant of n x n matrix consisting of $\epsilon$ in the diagonal elements and $\upsilon$ in the upper and lower sub-diagonal and the rest of the elements of the matrix being zero. }

\begin{equation}
\left( \begin{array}{l}
d_{n} \\
d_{n-1} \\
\end{array} \right) =
\left( \begin{array}{ll}
\epsilon & -\upsilon^{2} \\
1 & 0 \\
\end{array} \right) \left( \begin{array}{l}
d_{n-1} \\
d_{n-2} \\
\end{array} \right) 
\end{equation}

\vspace{0.2in}
\hspace{0.2in}

{\small By repeated application of the above recursion relation, and noting that $d_{0} = 1$ and $d_{1} = \epsilon$ one arrives at the following result}

\begin{equation}
d_{n} = \frac{\lambda_{1}^{n+1} - \lambda_{2}^{n+1}}{\lambda_{1} - \lambda_{2}}
\end{equation}

\begin{equation}
\lambda_{1,2} = \frac{\epsilon \pm \sqrt{\epsilon^{2} - 4 \upsilon^{2}}}{2}
\end{equation}

\vspace{0.2in}
\hspace{0.2in}

{\small Employing this result in the present Hamiltonian, yields the below given result for the determinant of $\tilde{\mathrm{H}}_{chain}$ }

\begin{equation}
\mathrm{det \lbrack \tilde{H}}_{chain} \rbrack_{1n} = d_{n} + i \Delta d_{n-1} 
\end{equation}

\vspace{0.2in}
\hspace{0.2in}

{\small Where in the above expression $d_{n}$ is the same as the one defined above . Care should be taken to note that the $\epsilon$ used in the definition of $\lambda_{1,2}$ is to be replaced with $\epsilon - \epsilon_{i}$. This is done since the quantity of interest is the inverse of $\epsilon \mathrm{I} - \tilde{\mathrm{H}}_{chain}$}

\vspace{0.2in}
\hspace{0.2in}

{\small In the above expression the green's function element obtained was of the isolated chain Hamiltonian, but the green's function element to be employed in the formalism should correspond to the full total Hamiltonian. This can be obtained from the isolated chain Hamiltonian's green's function by resorting to a Dyson equation with the interaction of the electrode and redox treated as a perturbation. The resulting expression is as follows:}

\begin{equation}
\mathrm{G}_{1N} = \frac{\mathrm{G}_{1N}^{0}}{\lbrack 1 - \mathrm{G}_{NN}^{0} \bar{\upsilon}_{nr}  ^{2}\mathrm{G}_{rr}^{0} \rbrack}
\end{equation}

\vspace{0.2in}
\hspace{0.2in}

{\small Where in the above expression the isolated green's function elements are given by}

\begin{equation}
\mathrm{G}_{NN}^{0} = \frac{d_{n-1} + i \Delta d_{n-2}}{d_{n} + i \Delta d_{n-1}}
\end{equation}

\begin{equation}
\mathrm{G}_{rr}^{0} = \frac{1}{\epsilon - \epsilon_{r} - \Sigma_{\nu} \hbar \omega_{\nu} g_{\nu} q_{\nu} \pm i \delta} 
\end{equation}

\vspace{0.2in}
\hspace{0.2in}

{\small Alternatively one can arrive at the expression for the $\mathrm{G}_{1n}$ element by considering only the Hamiltonian for the chain, $\mathrm{H}_{chain}$ and replacing the effect of the reservoir and redox by the corresponding self-energies.Working within the Wide-Band approximation where the self-energy of the electrode is given by ( $-i\Delta$ ). Both the procedures leads to the expression for $\mathrm{G}_{1n}$ as}

\begin{equation}
\mathrm{G}_{1n} = \frac{(-\upsilon)^{n-1}}{(d_{n} + i\Delta d_{n-1}) - \frac{\lbrack d_{n-1}  + i\Delta d_{n-2} \rbrack \bar{\upsilon}_{nr} ^{2}}{\epsilon - \epsilon_{r} - \Sigma_{\nu} \hbar \omega_{\nu} g_{\nu} q_{\nu}}}
\end{equation} 

\vspace{0.2in}
\hspace{0.2in}
 
{\small Where in the above expression wide-band approximation has been employed for the self-energy of the reservoir and the self-energy of the redox is assumed to first approximation as $\frac{\bar{\upsilon}^{2}}{\epsilon - \epsilon_{r} - \Sigma_{\nu} \hbar \omega_{\nu} g_{\nu} q_{\nu}}$}

\vspace{0.2in}
\hspace{0.2in}

{\small The above expression for $\mathrm{G}_{1n}$ is to be substituted in the rate expression. The delta function can also be split and hence the rate expression takes the form as shown below}

\begin{equation}
\kappa(\epsilon,q_{\nu}) = \frac{2 \pi }{\hbar} \int d\grave{\epsilon} \Sigma_{k} \lbrack 1 - f(\epsilon - e W) \rbrack \mid \bar{\upsilon}_{k1} \mid ^{2} \delta(\grave{\epsilon} - \epsilon_{k}) \mid \bar{\upsilon}_{nr} \mid ^{2} \mid \mathrm{G}_{1r} \mid ^{2} \delta (\epsilon_{r} + \Sigma_{\nu} \hbar g_{\nu} q_{\nu} - \grave{\epsilon}) 
\end{equation}

\vspace{0.2in}
\hspace{0.2in}

{\small Defining $\pi \mid \bar{\upsilon}_{k1} \mid ^{2} \Sigma_{k} \delta (\epsilon - \epsilon_{k} ) = \Delta $ the above expression reduces to the form below}

\vspace{0.2in}
\hspace{0.2in}

\begin{equation}
\kappa(\epsilon,q_{\nu}) = \frac{2 }{\hbar} \Delta \lbrack 1 - f(\epsilon - eW) \rbrack \mid \mathrm{G}_{1n} \mid ^2 \mid \bar{\upsilon}_{nr} \mid ^{2} 
\end{equation}

\vspace{0.2in}
\hspace{0.2in}

{\small The $\epsilon$ in the above expression refers to the energy at which the electron transfer takes place and hence to get the net current flow  it is required to sum all the contributions from different energy windows. This summing of different contributions can be performed by integrating with respect to $\epsilon$. That is a sum over $\epsilon$ is replaced with $\int d\epsilon \rho(\epsilon_{f})$ where $\rho(\epsilon_{f})$ is the density of states at the fermi-surface of the electrode .  }

\begin{equation}
\kappa (q_{\nu}) = \frac{2 }{\hbar} \Delta \int d\epsilon \rho(\epsilon_{f})  \lbrack 1 -f(\epsilon - eW) \rbrack \mid \mathrm{G}_{1n} \mid ^2 \mid \bar{\upsilon}_{nr} \mid ^2 
\end{equation}

\begin{equation}
\kappa (q_{\nu}) = \frac{2 }{\hbar} \Delta \rho(\epsilon_{f}) \int d\epsilon  \lbrack 1 -f(\epsilon - eW) \rbrack \mid \mathrm{G}_{1n} \mid ^2 \mid \bar{\upsilon}_{nr} \mid ^2
\end{equation}

\vspace{0.2in}
\hspace{0.2in}

{\small It is now required to do thermal averaging which is to integrate the above expression for current over all the polarization modes ,$q_{\nu}$, with a weighing factor of $exp^{-\beta E}$. Alternatively the same process can be viewed as summing over all possible initial states of the redox with respect to polarization modes with a suitable weighing factor depending on the energy of the polarization modes. }

\vspace{0.2in}
\hspace{0.2in}

{\small The above expression can be evaluated by employing the single reaction co-ordinated $Q$ for the polarization modes $q_{\nu}$. Resorting to single which in effect can be represented as a change of co-ordinates, the crucial observation is that the introduction of single reaction co-ordinate amounts to replacing $\Sigma_{\nu} \hbar \omega_{\nu} g_{\nu} q_{\nu} $ by $2\lambda q$ and the quadratic term in the exponential for the thermal average is to be replaced by $\lambda q^{2} + 2 \lambda q$. Thus the expression for net current takes the form as shown below}

\begin{equation}
\kappa = \frac{1}{\mathrm Z} \frac{2}{\hbar} \Delta \rho(\epsilon_{f}) \bar{\upsilon}_{nr} \mid^{2} \int d\epsilon \lbrack 1 - f(\epsilon - eW) \rbrack \int dq e^{-\beta(\lambda q^{2} + 2 \lambda q)} \mid \mathrm{G}_{1n} (q)\mid^{2} \mid 
\end{equation}

\begin{equation}
\mathrm{G}_{1n} (q) =  \frac{(-\upsilon)^{n-1}}{(d_{n} + i\Delta d_{n-1}) - \frac{\lbrack d_{n-1}  + i\Delta d_{n-2} \rbrack \bar{\upsilon}_{nr} ^{2}}{\epsilon - \epsilon_{r} +2 \lambda q}}
\end{equation}

\begin{equation}
 {\mathrm Z} = \int dq e^{-\beta(\lambda q^{2} + 2 \lambda q)} 
\end{equation}

\vspace{0.2in}
\hspace{0.2in}

{\small While the partition function Z is an elementary gaussian integral. The innermost $dq$ integral can be evaluated by using partial fractions. For brevity defining ${\mathrm D}_{n} = d_{n} + i \Delta d_{n-1}$, the ${\mathrm G}_{1n}$ takes a form as below. }

\begin{equation}
 {\mathrm G}_{1n} = \frac{(-\upsilon)^{n-1} (\epsilon - \epsilon_{r} + 2 \lambda q)}{D_{n} (\epsilon - \epsilon_{r} + 2 \lambda q) - \bar{\upsilon}^{2} D_{n-1}}
\end{equation}

\vspace{0.2in}
\hspace{0.2in}

{\small The innermost integral over $dq$ can be explictly evaluated}

\begin{eqnarray}
 \int dq e^{-\beta(\lambda q^{2} + 2 \lambda q)} \mid {\mathrm G}_{1n} \mid ^{2}  & = & \frac{(\upsilon)^{2n-2}}{2 \lambda} \lbrack {\mathrm A} e^{-\beta({\mathrm Q}_{0} + 2 \lambda^{2})/4 \lambda} (\epsilon - \epsilon_{r} + {\mathrm Q}_{0})^{2} \nonumber \\
& & + {\mathrm B} e^{-\beta(\bar{{\mathrm Q}}_{0}  + 2 \lambda^{2})/4 \lambda} (\epsilon - \epsilon_{r} + \bar{{\mathrm Q}}_{0})^{2} \rbrack
\end{eqnarray}

\vspace{0.2in}
\hspace{0.2in}

{\small where ${\mathrm A} = \frac{D_{n}}{\bar{\upsilon}_{nr} ^{2} (D_{n-1} \bar{D}_{n} -  \bar{D}_{n-1}   D_{n} ) } $ and ${\mathrm B}  = \frac{-\bar{D}_{n}}{\bar{\upsilon}_{nr} ^{2} (D_{n-1} \bar{D}_{n} -  \bar{D}_{n-1}   D_{n} ) }$ and $\bar{\mathrm Q}_{0}$ are complex conjugates of each other. Physically this corresponds to the pole in ${\mathrm G}_{1n} (q)$.    }

\begin{equation}
 {\mathrm Q}_{0} = \bar{\upsilon}_{nr} ^{2} \frac{D_{n-1}}{D_{n}} - (\epsilon - \epsilon_{r})
\end{equation}

\begin{equation}
 {\mathrm A} = \frac{D_{n}}{2 i \Delta (d_{n-2} d_{n} - d_{n-1} d_{n-1})}
\end{equation}

\begin{equation}
 {\mathrm B} = \frac{-\bar{D}_{n}}{2 i \Delta (d_{n-2} d_{n} - d_{n-1} d_{n-1})}
\end{equation}

\begin{equation}
 \kappa = {\mathrm P.F}  \int d\epsilon  \frac{1 - f(\epsilon - eW)}{2 i \Delta (d_{n-2} d_{n} - d_{n-1} d_{n-1})} \lbrack D_{n} e^{-\beta(\bar{{\mathrm Q}}_{0}  + 2 \lambda) ^{2} /4 \lambda} (\bar{\upsilon}_{nr} ^{2} \frac{D_{n-1}}{D_{n}} )^{2} - c.c \rbrack
\end{equation}

\begin{equation}
 {\mathrm P.F} = \frac{2}{\hbar} \frac{\sqrt{\beta}}{\sqrt{\pi \lambda}} \Delta \mid \bar{\upsilon}_{nr} \mid ^{2} \frac{(\upsilon)^{2n-2}}{2 }  e^{-\beta \lambda}
\end{equation}

\vspace{0.2in}
\hspace{0.2in}

{\small The above expression for current can be written in a simplified form by resorting to a step function approximation to $1 - f(\epsilon - eW)$ .}

\begin{equation}
 \kappa = {\mathrm P.F} \int _{\epsilon_{f} -W} ^{+\infty} \frac{\lbrack D_{n} e^{-\beta(\bar{{\mathrm Q}}_{0}  + 2 \lambda) ^{2} /4 \lambda} (\bar{\upsilon}_{nr} ^{2} \frac{D_{n-1}}{D_{n}} )^{2} - c.c \rbrack}{2 i \Delta (d_{n-2} d_{n} - d_{n-1} d_{n-1})} d\epsilon
\end{equation}

\vspace{0.2in}
\hspace{0.2in}

\section{Results and Discussion}

\vspace{0.2in}
\hspace{0.2in}

{\small Our main focus of attention in this work is oriented towards explaining the dynamic interaction of the redox with the solvent modes and it's subsequent effect on the transfer rate. The chief reason being the redox energy levels varies with the $q$ and hence even though maxima in the transfer rate is expected when the energy level of the redox is same as the energy level of the bridge, but due to the coupling of the redox with the solvent modes, the value of $q$ coupled with the  resonant energy level maynot be a  thermally favoured one. Thus it's the competition between the energy difference between the redox and the bridge and the probability distribution of the initial state of the redox that brings out certain interesting results which we focus on this part. }

\vspace{0.2in}
\hspace{0.2in}

\begin{figure}
 \begin{center}
  \includegraphics[scale=0.6,angle=-90]{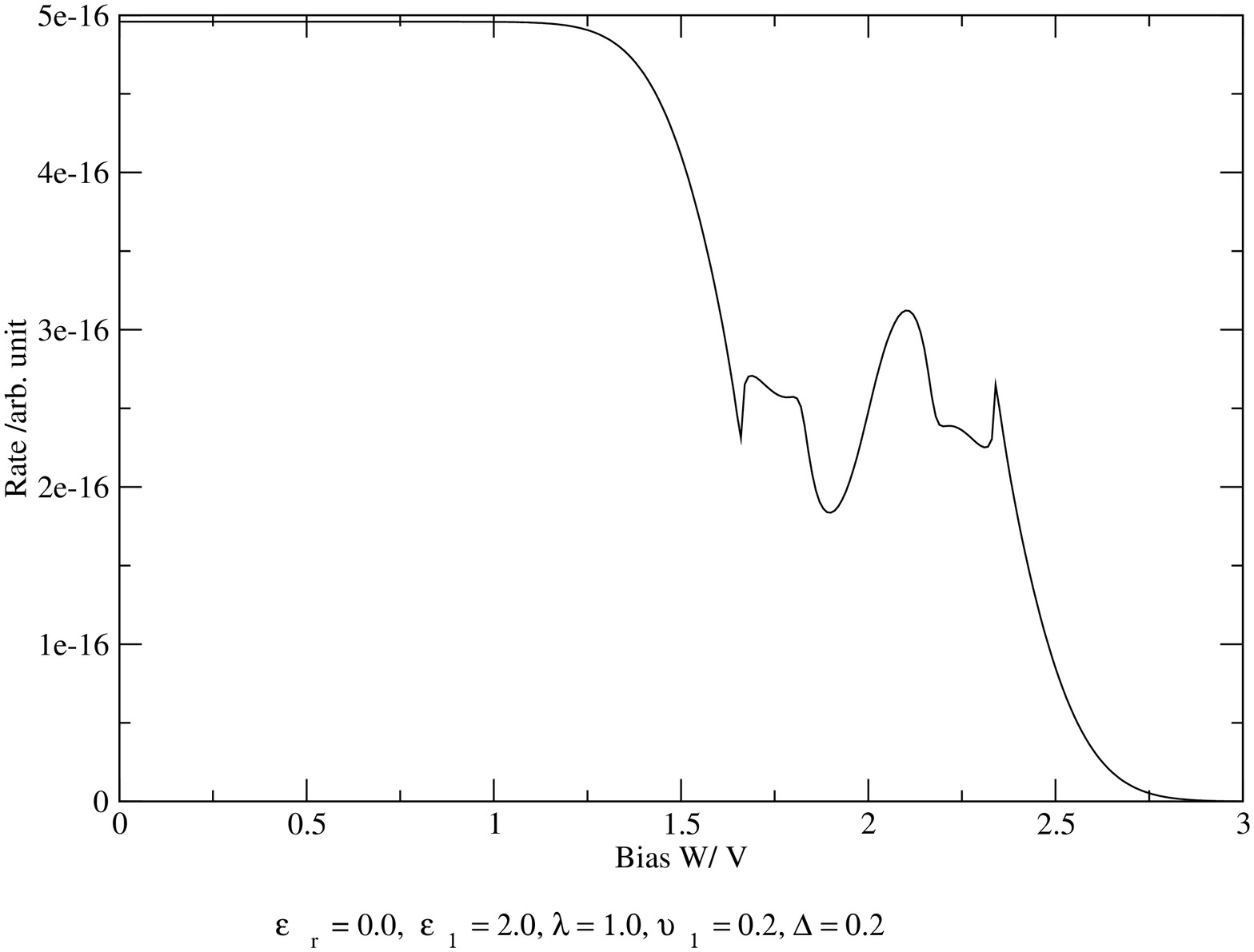}
\label{rvr1}
 \end{center}
\end{figure}

{\small Since there exist a number of parameters as input, unless stated specifically, we set the fermi surface to zero ,$\epsilon_{f} = 0$, $\Delta = 0.2$ and $\bar{\upsilon}_{nr} = \upsilon_{i}$, for all through the rest of the discussion. Fig ~\ref{rv1}. shows the variation of the rate with the applied voltage which could be understood easily from a naive argument as follows: As the applied voltage is increased, the energy levels available in the electrode for the electrons originating from the redox decreases. Thus leading to a rate approaching zero as the voltage is increased. The above is a normally expected behaviour while the interplay between the polarisation modes and the resonance between the redox and the bridge energy levels is clearly visible in Fig ~\ref{rvr1}. Analysis reveals that the mode corresponding to $q = -1$ is the thermally favoured mode, and hence when the energy difference between redox and bridge sites equals $2 \lambda$ is the condition for energy matching between the redox and bridge. This condition is satisfied for the set of values employed in obtaining Fig ~\ref{rvr1}. For the set of values used, the resonance between redox and bridge occurs at 2 eV. Thus patterns with local maximas and minimas occur around 2 eV in Fig ~\ref{rvr1} while the curve falls smoothly and rapidly around the same value in Fig ~\ref{rvr1} .}

\vspace{0.2in}
\hspace{0.2in}

{\small Fig ~\ref{rvt}. is plotted to show the temperature variation of the rate at zero bias. As expected, the graph shows a linear behaviour when plotted as a function of (1/T). Taking logarithm of the rate equation reveals only three terms dependent on $\beta$, apart from uninteresting constant terms. One of the terms is -0.5 log $\beta$ while the other two terms have linear dependency on $\beta$. The contribution from log $\beta$ become dominant only in the $\beta \rightarrow 0$ limit. Since this corresponds to the high temperature limit , for all practical purposes the log variation of rate has a linear dependency on (1/T), a fact borne by Fig ~\ref{rvt}. }

\vspace{0.2in}
\hspace{0.2in}

\begin{figure}
 \begin{center}
  \includegraphics[scale=0.6,angle=-90]{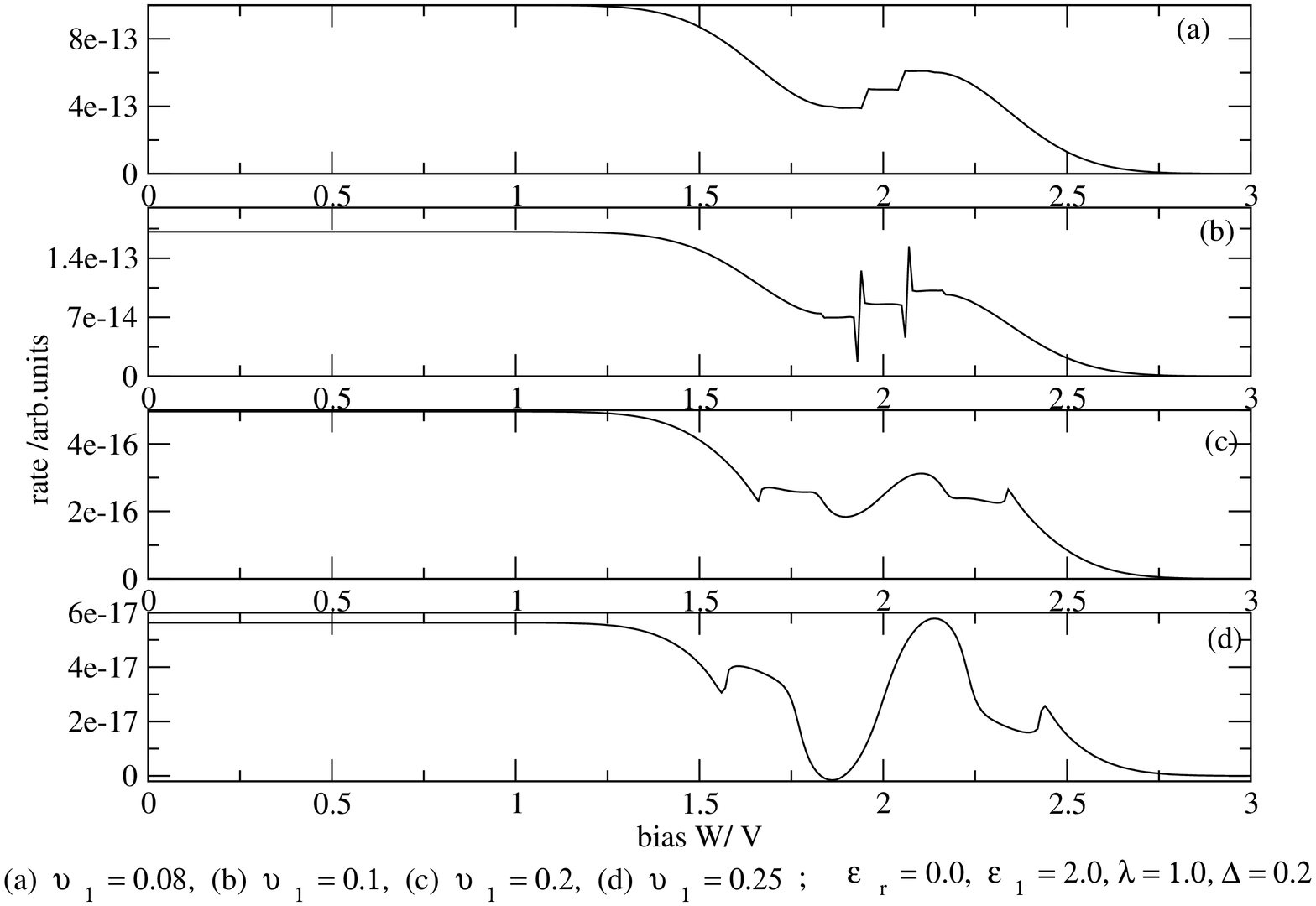}
\label{rvvb}
 \end{center}
\end{figure}

{\small In order to further investigate the properties of transfer rate at resonance condition, we plot in Fig ~\ref{rvvb}. the response of the rate-Voltage for various values of $\upsilon_{1}$. It is observed from the plots that for small values of the coupling co-effecients , implying a weak bonding between the bridge atoms and between the bridge atoms and the redox, drastic variation from the normal behaviour is not seen. But as the value of $\upsilon_{1}$ is varied, the observed pattern near the resonance point changes. The changes include a constant transfer rate around a small range of voltage near the voltage value corresponding to the resonance energy, to the rate dropping to zero in some intermediate voltage regime.}

\vspace{0.2in}
\hspace{0.2in}

\begin{figure}
 \begin{center}
  \includegraphics[scale=0.6,angle=-90]{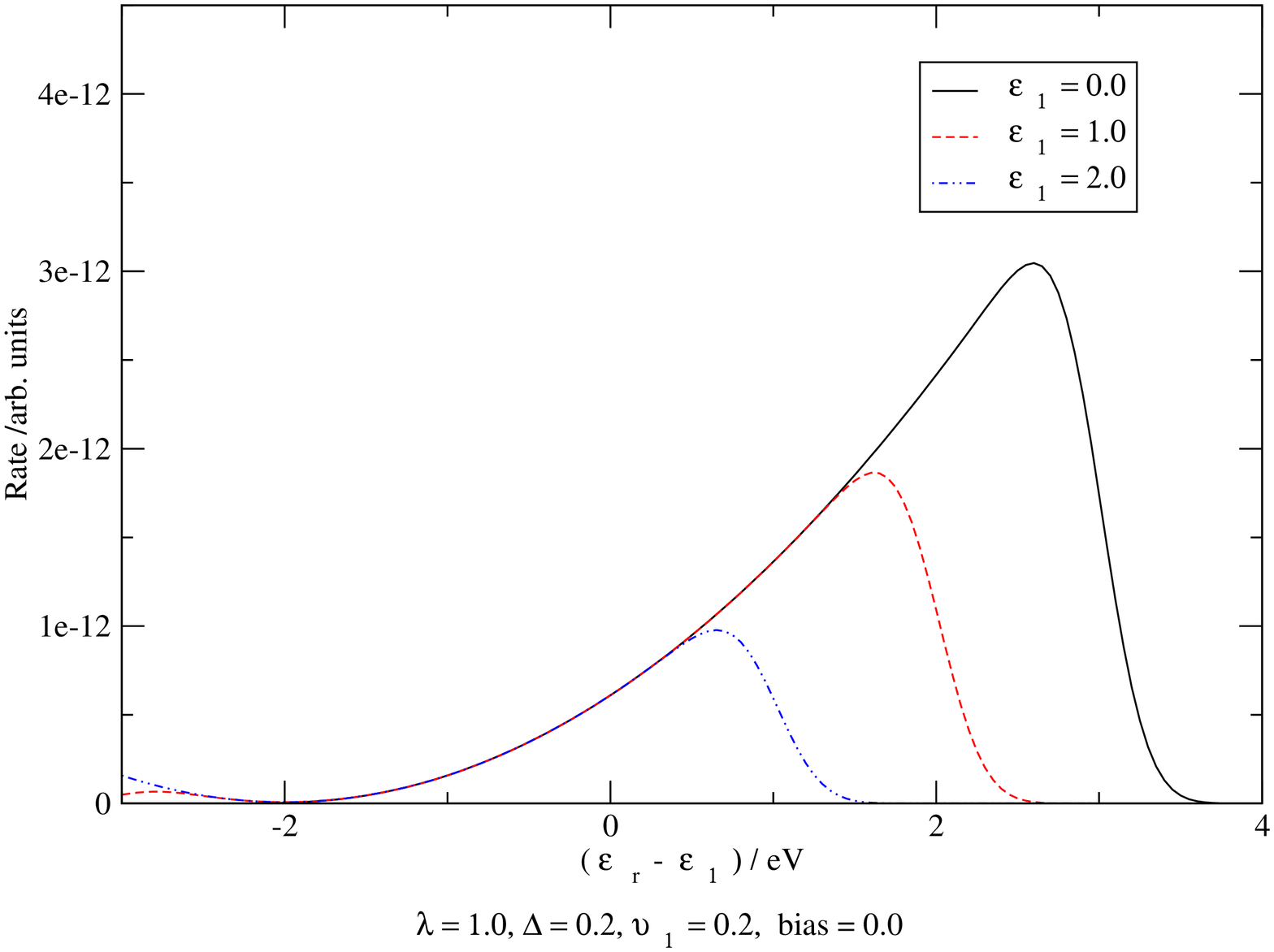}
\label{rvde}
 \end{center}
\end{figure}

\begin{figure}
 \begin{center}
  \includegraphics[scale=0.6,angle=-90]{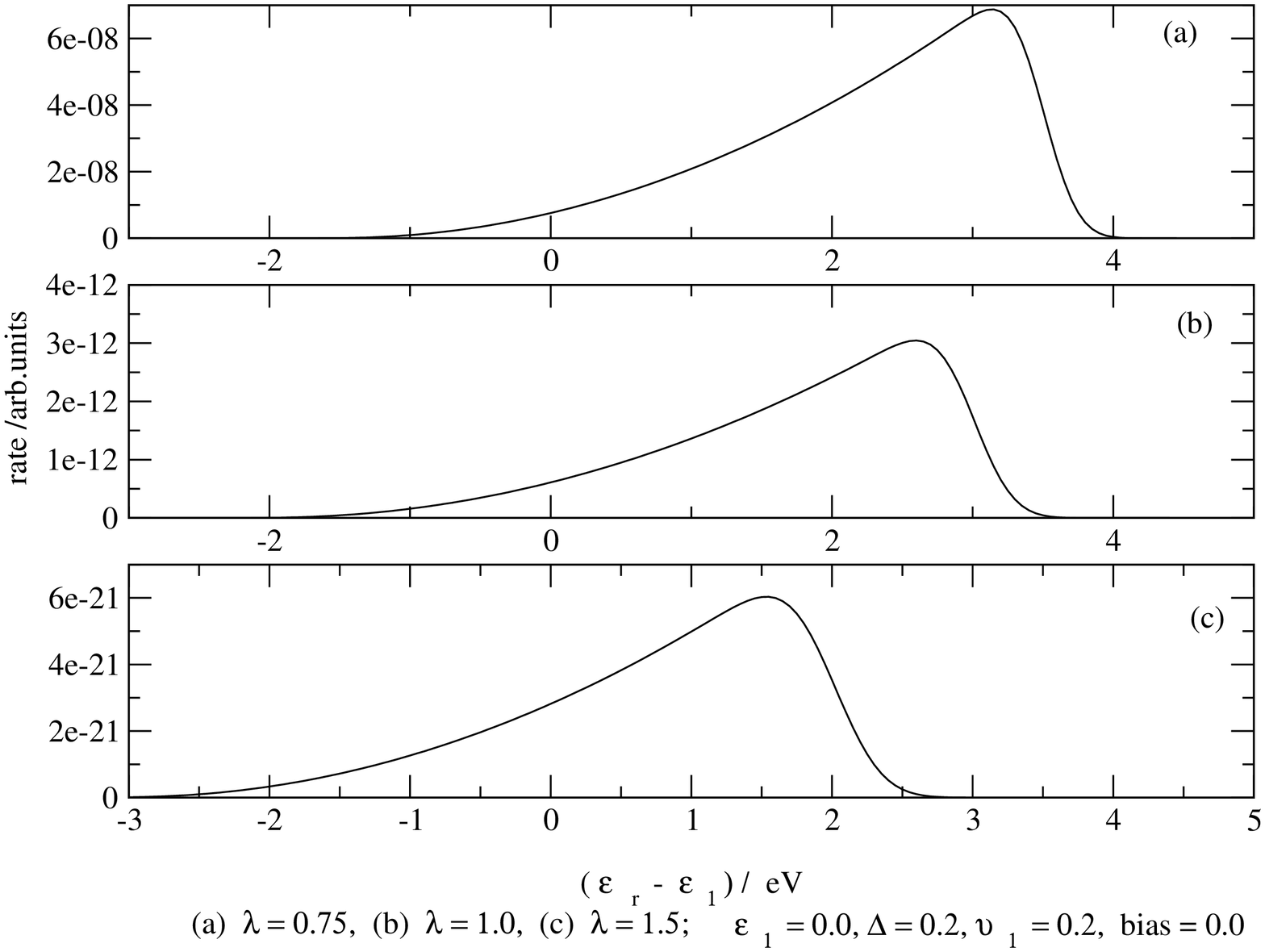}
\label{rvdl}
 \end{center}
\end{figure}

{\small In Fig ~\ref{rvde}. we plot the variation of the rate as a function of the difference in energy between the bridge and the unsolvated redox energy. In particular we wish to highlight the behaviour in the regime where the bridge is situated at a higher energy level than the fermi level. we first study the variation for a fixed value of $\lambda$. The crucial fact to observe is that as long as the bridge is located energetically higher than the fermi surface,  the location of the maximas varies with the actual location of energy level of the bridge and not just only upon the energy difference between the redox and the bridge as one would normally expect.   On examining the plots show that the maxima occurs for a fixed value of $\epsilon_r$ corresponding to a particular value of $\lambda$. Thus in effect while the minima is dependent on the difference between the energy levels, the maxima shows a strong dependecy on the solvated energy level of the redox. The effect of $\lambda$ on the rate profile is shown in Fig ~\ref{rvdl}. }

\vspace{0.2in}
\hspace{0.2in}

\section{conclusion}

\vspace{0.2in}
\hspace{0.2in}

{\small In this article we have considered a model for electron transfer rate from redox to an electrode via a bridge of N- atoms. The redox considered is allowed to interact with an electrochemical system whose effect in terms of polarization modes are modelled as a bath of classical oscillators with solvent co-ordinates $q$. We obtained the electron transfer rate from redox to the electrode and have performed an averaging over all possible initial states of the redox. A final voltage dependent expression for the rate is obtained as an integral over all possible energy values available for the incoming electron. This integral has been performed numerically. }

\vspace{0.2in}
\hspace{0.2in}

{\small We have shown some of the interesting pattern which are otherwise not expected in a heuristic analysis. These pattern in the voltage dependency have been attributed to the competing process namely,resonance tunneling between the redox and the bridge and the solvation of the redox. Additionally we have also shown that by suitable altering the voltage range and the coupling co-efficients it's possible to block the electron transfer in some intermediate voltage ranges. Moreover, the profile of the transfer rate with respect to the energy difference between the bridge and the redox exposes a radically different response due to the effect of intervening solvent modes. }

\newpage
\subsection*{Figures}

\begin{figure}
 \begin{center}
  \includegraphics[scale=0.6,angle=0]{rate_V_1.eps}
\label{rv1}
 \end{center}
\end{figure}

\begin{figure}
 \begin{center}
  \includegraphics[scale=0.6,angle=0]{Rate_vs_T.eps}
\label{rvt}
 \end{center}
\end{figure}

\end{document}